\documentclass[aps,twocolumn,amsfonts,amssymb,longbibliography,superscriptaddress]{revtex4-1}
\usepackage{graphicx}
\usepackage{epstopdf} 
\usepackage{dcolumn}
\usepackage{bm}
\usepackage{bbm,bbold}
\usepackage{color}
\usepackage{xcolor, soul}
\sethlcolor{green}
\usepackage{amsfonts}
\usepackage{amsmath}
\usepackage{mdframed}
\usepackage[normalem]{ulem}
\usepackage{mathrsfs}   
\usepackage[none]{hyphenat}
\usepackage{subfigure}
\usepackage{float}
\usepackage{physics}
\usepackage[colorlinks=true,citecolor=blue]{hyperref}
\hypersetup{colorlinks=true,citecolor=blue,linkcolor=blue,urlcolor=blue}

\begin{document}
	\title{Tunable optical bistability of two-dimensional tilted Dirac system}
	\author{Vivek Pandey}
	\affiliation{Department of Physics, School of Engineering and Sciences, SRM University AP, Amaravati, 522240, India}
	\author{Pankaj Bhalla}
	\affiliation{Department of Physics, School of Engineering and Sciences, SRM University AP, Amaravati, 522240, India}

\date{\today}
\begin{abstract}
We study the phenomenon of controlling the light by light known as the optical bistability for the two-dimensional tilted Dirac system. Using the Boltzmann approach under relaxation time approximation, we find that the optical bistability can be controlled by the nonlinear response of the system. For the prototype, we consider an inversion symmetry broken system. We find that the optical bistability associated with the nonlinear response is tunable with the strength of the tilt, gap and chemical potential. This suggests the inputs for the development of future-generation optical devices. 
\end{abstract}

\maketitle

\section{Introduction}
Investigations of the nonlinear responses of two-dimensional materials to the light beams have geared up in recent years due to immense applications in optoelectronics like optical switching~\cite{dai_SR2015, mazurenko_PRL2003}, signal processing~\cite{yan_JLT2012}, optical transistors~\cite{Abraham_1982RPP}, modulators~\cite{nurrohammadi_Optik2022}, slow light devices~\cite{nurrohammadi_MSSP2022} and logic gates~\cite{assanto_APL1995,smith_Nat1985}. In particular, research has focused on photovoltaic~\cite{chan_PRB2017}, second-harmonic~\cite{you_NP2019, Wang_OME2019, khan_AFM2022}, third-harmonic~\cite{Cheng_2019,shi_molecules2023}, circular photogalvanic effects~\cite{aftab_NS2023}, etc for time-reversal and inversion symmetric systems~\cite{huang_PRB2023}. Moreover, the third-order nonlinear optical effects have also been studied in nanostructured systems using the two-wave mixing process considering the first and second harmonic beams~\cite{Hernández-Acosta_2019}. The salient example among such systems is graphene, a time-reversal and inversion symmetric system~\cite{bona_NP2010}. It has a linear band dispersion relation and band touching at the Dirac point in the Brillouin zone which allows the interband transitions in the wide frequency range~\cite{neto_RMP2009}. The present centrosymmetric structure of the system forbids the second harmonic generation effect~\cite{Orenstein_ARCM2021, boyd_book, shen_book}. However, the restriction can be lifted by breaking the inversion symmetry by incorporating the tilt in the system~\cite{Gao_OE2021, Guo_eL2021}. This includes tilted graphene, Dirac semimetals having band touching points with opposite chirality~\cite{mojarro_PRB2021, xiong_PRB20023}, Weyl semimetals~\cite{trescher_PRB2017, Mukherjee_JPCM2018, Shao_JPCM2023, das_PRB2019} having nondegenerate band touching points, etc.~\cite{armitage_RMP2018, yan_ARCMP2017,hu_ARMR2019}. Thus, the tilted Dirac systems, which permit all effects, are drawing ever-growing experimental and theoretical interest~\cite{balle_PRB2023, ware_PRB2023, ong_NRP2021, Kundu_NJP2020}.

Optical bistability is a nonlinear phenomenon where a system possesses two stable output values or transmitted intensities for a given input value or incident intensity~\cite{gibbs_book}. Thus, the system can be rapidly transformed between two states, which refers to switching. Traditionally, optical bistability is configured with the Fabry-Perot cavity filled with a nonlinear medium and achieved by the combined effect of two entities~\cite{gibbs_PRl1976, smith_APL2008}. One is the nonlinear medium, which responds uniquely to the field. Another is the feedback mechanism that is proportional to the output. It has been shown that the bistability depends on the large nonlinear effect which can be obtained by two methods. One is the modulation of the refractive index of the nonlinear material with light~\cite{goldstone_PRL1984, xiang_APL2014}. This requires the thick nonlinear medium to have a large nonlinear Kerr index. The second method is the field effect on the material in nonlinear regimes. The ultimate obstacle is the choice of material and structure that requires low input power to enhance nonlinear optical effects for bistable devices~\cite{Abraham_RPP1982}. Thanks to graphene, which paves the way to realize optical bistable devices due to the sizeable nonlinear response and large Kerr effect~\cite{hendry_PRL2010, MIKHAILOV_PE2012, Mikhailov_JPCM2008, MIKHAILOV_ME2009, yao_PRL2012}. Later, it is extended with graphene–silicon resonator~\cite{kim_NP2012}, graphene nanoribbons intercalated with boron nitride~\cite{zhang_JMCC2014}, and graphene nanobubbles~\cite{bao_AOM2015}. Dai et al.~\cite{dai_SR2015} observed the optical bistability due to the excitation of graphene surface plasmons. Further, Optical bistability has been demonstrated with the modulation of the charge carrier density of graphene~\cite{Dai_OE2015}.

Theoretically, the nonlinear responses have been discussed within the traditional Boltzmann approach under relaxation time approximation by expanding the nonequilibrium distribution function in terms of the field~\cite{sodemann_PRL2015, zyuzin_PRB2018, morimoto_PRB2016} and quantum kinetic approaches~\cite{dai_CPR2023, sipe_PRB2000, bhalla_PRL2020, bhalla_PRL2022, juan_PRX2020, Ishizuka_2017}.
Recently, the Boltzmann approach has also been used to investigate various effects such as the dynamically tunable optical bistability in three-dimensional system by exciting topological excited states using the photonic crystals in the terahertz region~\cite{long_OE2022}, nonlinear plasmonic effects in three-dimensional Dirac semimetals~\cite{Ooi_APLP2018}, nonlinear magnetotransport in two-dimensional system~\cite{pal_PRB2023}, electro-optic response in two-dimensional materials~\cite{Rappoport_PRL2023}, nonlinear optical response in two-dimensional semi-Dirac system~\cite{Dai_JPCM2019}. 
However, there are tiny studies that deal with the optical bistability phenomenon. Due to the strong nonlinearity in graphene, Peres et al. discuss the realization of the optical bistability of graphene in the high-frequency regime~\cite{peres_PRB2014}. However, the studies uncover the interesting questions. (1) Why have tilt and gap terms been neglected so far? (2) What will be the consequences of such terms to the bistability effect? To have an in-depth understanding of these insightful questions, we analyze the optical bistability phenomenon for the two-dimensional tilted Dirac material. We show that the tilt does not alter the Berry curvature of the system but provides tunable optical bistability. Here we also study the effect of a gap around the Dirac point which provides a significant value to the response. Thus an extensive analysis of optical bistability with gap and tilt gives a better description, required before accessing tilted Dirac material for optical devices.

The paper is organized as follows. In Sec.~\ref{sec:methodology} we present a model Hamiltonian of the tilted Dirac system, and we present the theoretical approach to obtain the nonlinear distribution function. Later, we derive the nonlinear currents. In Sec.~\ref{sec:OB} we give the derivation of the optical bistability. In Sec.~\ref{sec:results} we analyze the bistability effect for the tilted Dirac system with the tilt, gap, incident frequency, and chemical potential. We summarize our work in Sec.~\ref{sec:summary}.
\begin{figure*}[htp]
    \centering
    \includegraphics[width=18cm]{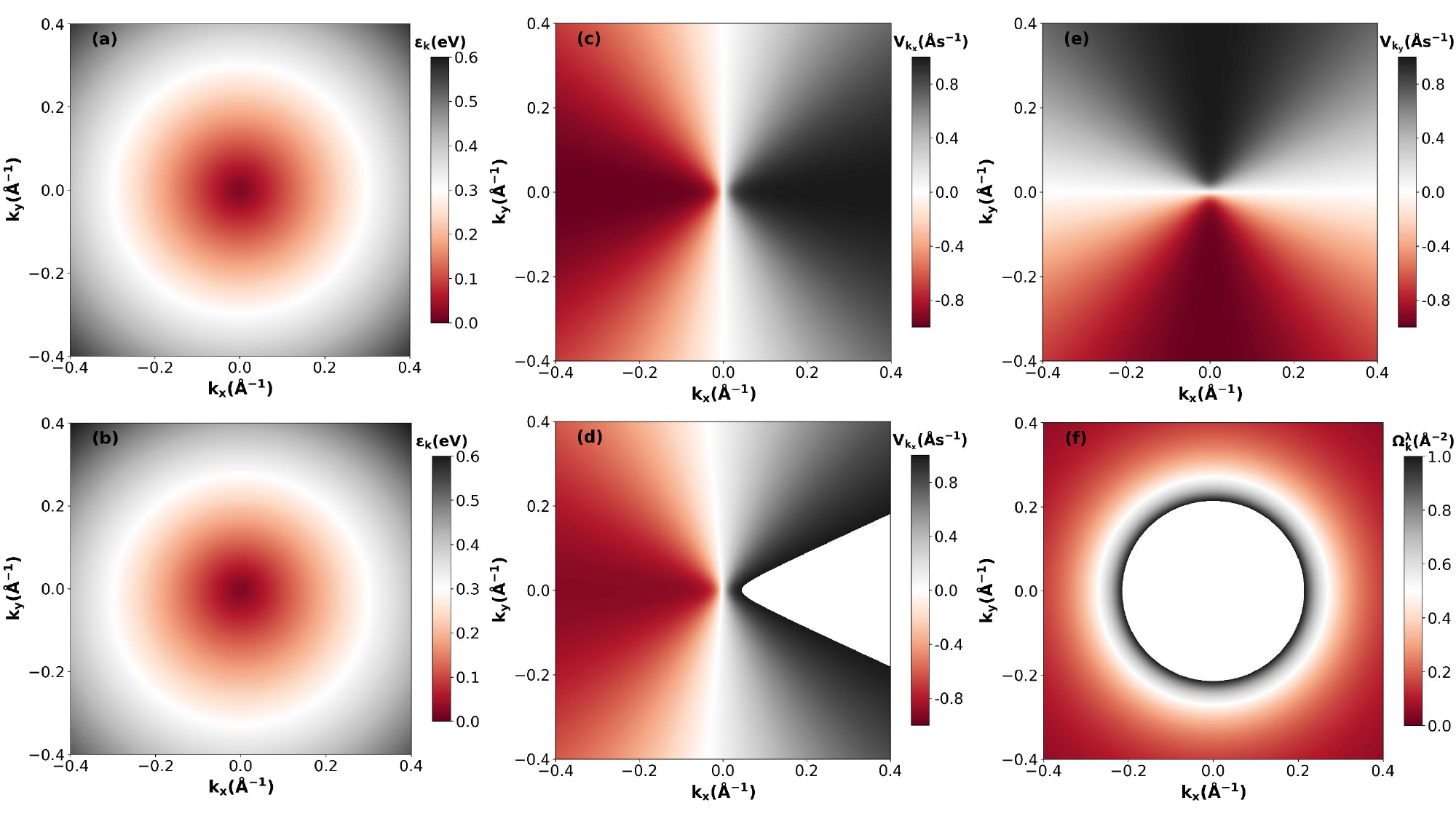}
    \caption{(a) Density plot for the energy dispersion for the tilted two-dimensional Dirac system at tilt $t=0$ and (b) at tilt $t=0.1$, (c) and (d) velocity along $x$-direction at $t=0$ and finite tilt, $t=0.1$ respectively, (e) velocity along $y$-direction and (f) Berry curvature for the tilted Dirac system. Here we set $\Delta = 0.02$ eV to illustrate the behavior of quantities in the momentum space and the color bars to show the strength of the respective quantities.}
    \label{fig:dispersion}
\end{figure*}
\section{Model and Method}
\label{sec:methodology}
We consider a ${\bf k} \cdot {\bf p}$ effective low energy model Hamiltonian for the two-dimensional tilted Dirac material, which has the following representation~\cite{sodemann_PRL2015, du_PRL2018}
\begin{equation}
    \mathcal{H}({\bm k}) =  \hbar t\zeta k_x \sigma_0 +  \hbar v_x \zeta k_x \sigma_y - \hbar v_y k_y \sigma_x + \Delta \sigma_z, 
\end{equation}
where $t$ is the parameter that introduces tilt in the band dispersion along the $k_x$-direction, $\zeta = \pm$ for time-reversal counterparts or the valley index, $v_i$ ($i \equiv x,y,z$) represents the velocity and $k_i$ corresponds to the wave vector of an electron in the $i^{\rm th}$-direction, $\sigma_i$'s are the Pauli matrices, and $\sigma_0$ is the $2 \times 2$ identity matrix. The parameter $\Delta$ refers to the gap in the Dirac spectrum. The energy eigenvalues for the corresponding Hamiltonian are
\begin{equation}
    \varepsilon_{\bm k} = \hbar t \zeta k_x + \lambda \sqrt{\hbar^2(v_x^2 k_x^2 + v_y^2 k_y^2) + \Delta^2}.
\end{equation}
Here $\lambda = +$ corresponds to the conduction band and $\lambda = -$ to the valence band. For the sake of simplicity, we set $v_x = v_y = v$ hereafter. The eigenvectors are
\begin{equation}
    | u_{\bm k}^{\pm} \rangle = \frac{1}{\sqrt{2}} \left(\begin{matrix}
     i \sqrt{1 \pm \Delta / \varepsilon_0} e^{-i\theta} \\
     \sqrt{1 \mp \Delta / \varepsilon_0}
    \end{matrix}\right),
\end{equation}
where $\varepsilon_0 = \sqrt{\hbar^2 v^2(k_x^2 + k_y^2) + \Delta^2}$ and $\theta$ is the polar angle obtained along the $k_x$-direction and is defined as $\theta = \tan^{-1} [k_y/\zeta k_x]$. Here the system without the tilt term preserves the time-reversal and inversion symmetry and follows the linear dispersion like graphene. However, the finite tilt breaks inversion symmetry and time-reversal symmetry for each valley due to the linear momentum factor. In addition, the mirror symmetry is also broken along the $k_x$-direction~\cite{nandy_PRB2019}. 

The velocity in the band basis is defined as 
\begin{align}
    v_{{\bm k}}^{\lambda} = \frac{1}{\hbar} \frac{\partial \varepsilon_{\bm k}^{\lambda}}{\partial {\bm k}} + {\bm \Omega}^{\lambda} \times {\bm E},
\end{align}
where the first term refers to the band velocity and the second term corresponds to the anomalous velocity with $\Omega^\lambda_{\bm k}$ as the Berry curvature and $\bm{E}$ is the electric field. Mathematically, the Berry curvature is defined as~\cite{xiao_RMP2010}
\begin{align}
    \Omega^\lambda_{\bm k} = \frac{2e}{\hbar}\epsilon_{abc} {\rm Im}\sum_{\lambda \neq \lambda'} \frac{\langle u_{\bm k}^{\lambda}\vert \partial \mathcal{H}/\partial k_b \vert u_{\bm k}^{\lambda'} \rangle \langle u_{\bm k}^{\lambda'} \vert \partial \mathcal{H}/\partial k_c \vert u_{\bm k}^{\lambda} \rangle}{(\varepsilon^\lambda_{\bm k} - \varepsilon^{\lambda'}_{\bm k})^2},
\end{align}
having $\epsilon_{abc}$ as Levi-Civita symbol and $(\varepsilon^\lambda_{\bm k} - \varepsilon^{\lambda'}_{\bm k})$ is the energy difference between energy corresponding to the bands $\lambda$ and $\lambda '$. Corresponding to the given Hamiltonian for a tilted Dirac system, the band velocity gives $\hbar t \zeta +\lambda  \hbar v^2 k_x/\varepsilon_0$ along the $\hat{x}$-direction and $\lambda \hbar v^2 k_y/\varepsilon_0$ along the $\hat{y}$-direction. The Berry curvature obtains~\cite{sodemann_PRL2015,du_PRL2018}

\begin{align}
    \Omega_{\pm}^{z,{\bm k}} = \mp \frac{e\zeta \Delta v^2 \hbar^2}{2\hbar \varepsilon_{0}^3}. 
\end{align}
The wave vector derivative of the Berry curvature is given in the form
\begin{align}
    \bigg(\frac{\partial \Omega_{\pm}^{z,{\bm k}}}{\partial k_x},\frac{\partial \Omega_{\pm}^{z,{\bm k}}}{\partial k_y}\bigg) &= \pm \frac{e}{\hbar} \frac{3\zeta \Delta v^4 \hbar^4}{\varepsilon_{0}^5} \big(2 k_x, 2 k_y\big).
\end{align}
Clearly, the presence of the tilt term does not affect the eigenvectors, hence the Berry curvature and the wave vector derivative of the Berry curvature of the system. For the considered model Hamiltonian, the corresponding dispersion, velocity, and the Berry curvature are shown in Fig.~\ref{fig:dispersion}. Note that without the tilt parameter and zero gap, the band dispersion is identical to the graphene. At large $\Delta$ as compared to the spin-orbit coupling term, it behaves as a massive fermion system~\cite{Chen_science2010}. Here, the gap between the conduction and valence band is $2\Delta$. In addition, the Berry curvature vanishes at zero gap value, but at a finite gap around the Dirac point, remains finite~\cite{sodemann_PRL2015, du_PRL2018}. Also, $\partial {\bm \Omega}_{\bm k} / \partial k_i$ with $i = (x,y$) behaves similarly to the Berry curvature. However, unlike the Berry curvature, it vanishes at origin $k_x = k_y = 0$.

\subsection{Boltzmann Approach}
We embark with the kinetic equation within the semi-classical Boltzmann transport approach for the non-equilibrium distribution function $f({\bm k},t)$, which reads~\cite{ashcroft_book, ziman_book}
\begin{align} \label{eqn:Boltz}
    \dfrac{\partial f({\bm k},t)}{\partial t} - \frac{e {\bm E}(t)}{\hbar} \cdot \dfrac{\partial f({\bm k},t)}{\partial {\bm k}} = - \frac{f({\bm k},t)- f_0({\bm k})}{\tau},
    \end{align}
where ${\bm E}(t) = E_0 e^{i\omega t} \hat{x}$ is a time-dependent and spatially homogeneous electric field along the $x$-direction having $E_0$ a field amplitude and $\omega$ as frequency, $\tau$ is the relaxation time scale which we considered as a constant parameter under the relaxation time approximation, $f_0({\bm k})$ is the equilibrium Fermi-Dirac distribution function and is defined as $f_0({\bm k}) = [1 + e^{-\beta (\varepsilon_{\bm k} - \mu)}]^{-1}$ having $\mu$ as a chemical potential, $\beta = 1/k_B T$ with $k_B$ the Boltzmann constant, and $T$ the electron temperature.

To solve the kinetic equation, we expand the distribution function in terms of the electric field as
\begin{align}
    f({\bm k},t) = f_0({\bm k},t) + f_1({\bm k},t) + f_2({\bm k},t) + f_3({\bm k},t) + \cdots
\end{align}
and plug into the Boltzmann transport equation Eq.~\eqref{eqn:Boltz}. This obtains
\begin{align} \label{eqn:f}
    \Dot{f_N} ({\bm k},t) + \frac{f_N({\bm k},t)}{\tau} = \frac{e}{\hbar} {\bm E}(t) \cdot {\bm \nabla}_{\bm k} f_{N-1}({\bm k}),
\end{align}
where $\Dot{f_N} ({\bm k},t) = \partial f_N({\bm k},t)/\partial t$. Further on solving Eq.~\eqref{eqn:f} with the integrating factor, the $N^{\text{th}}$-order solution becomes
\begin{align} 
\label{eqn:Nthorder}
%
    f_N({\bm k},t) &= \frac{e}{\hbar} {\bm E_0} \cdot {\bm \nabla}_{\bm k} f_{N-1}({\bm k}) \frac{e^{-i\omega t}}{1/\tau - i\omega}   + c.c.
\end{align}
From the above equation, it is clear that the behavior of the distribution function corresponding to the particular field depends on the knowledge of the previous order distribution function. Here, for $N=1$, the factor $f_{N-1}({\bm k})$ on the right-hand side of the equation approaches the equilibrium distribution function $f_0$. For $N=2$, the $f_2$ corresponding to the second power of the electric field is given by
\begin{align}
    f_2({\bm k},t) &= e^{-t/\tau} \int_{-\infty}^{t} dt' e^{t'/\tau}e^{-i\omega t'}  \frac{e}{\hbar} {\bm E_0} \cdot {\bm \nabla}_{\bm k} f_1({\bm k},t) + c.c.
\end{align}
Substituting the expression for $f_1({\bm k},t)$, we get
    %
    %
%
\begin{align} \nonumber
    f_2({\bm k},t) &= \bigg\{ \frac{e^{-2i\omega t}}{(1/\tau - i\omega)(1/\tau - i 2\omega)} + \frac{\tau}{1/\tau - i\omega}  \bigg\}\\[2ex]
    &\times \frac{e^2}{\hbar^2} {\bm E_0} \cdot {\bm \nabla}_{\bm k} \bigg\{ {\bm E_0} \cdot {\bm \nabla}_{\bm k} f_0({\bm k}) \bigg\} + c.c.
\end{align}
The above expression comprises two terms due to the addition  and the cancellation of the exponential factor $e^{i\omega t}$ which lead to the second-harmonic and optical rectification effects. 

Similarly, the third-order distribution function $f_3$ with $N=3$ in Eq.~\eqref{eqn:Nthorder} yields
%
%
%
\begin{align} \nonumber
    & f_3({\bm k},t) = \bigg\{ \frac{e^{-3i\omega t}}{(1/\tau - i\omega)(1/\tau - i 2\omega)(1/\tau - i 3\omega)}\\[2ex] \nonumber
     & + \frac{e^{-i\omega t}}{(1/\tau - i\omega)^2(1/\tau - i 2\omega)}
    + \frac{2 e^{-i\omega t}}{(1/\tau - i\omega)^2 (1/\tau + i\omega)}  \bigg\}\\
    & \times \frac{e^3}{\hbar^3} {\bm E_0} \cdot {\bm \nabla}_{\bm k} \bigg\{ {\bm E_0} \cdot {\bm \nabla}_{\bm k} \bigg\} \bigg[ {\bm E_0} \cdot {\bm \nabla}_{\bm k} f_0({\bm k}) \bigg] + c.c.
\end{align}
Clearly, $f_3$ plays a significant role in generating the third harmonic and Kerr effects. 
\subsection{Currents}
The electric current in response to the applied electric field is defined as the trace of the product of the Bloch velocity and the non-equilibrium distribution function~\cite{ziman_book}. Mathematically, one can write like ${\bm j}(t) = -e{\rm{Tr}} [{\bm v}_{{\bm k}} f({\bm k},t)]$. For the first-order electric field, the linear current becomes
\begin{align}
    j_x^{(1)}(t) = -e^2 E_0^x \sum_{\bm k} \sum_{\lambda} (v_{\bm k}^{\lambda})^2 \pdv{f_0}{\varepsilon_{\bm k}} \frac{e^{-i\omega t}}{1/\tau - i\omega}. 
\end{align}
Here, the summation over $\lambda$ refers to the band summation and we replace the momentum derivative of the Fermi distribution function as $\partial f_0 / \partial {\bm k}= \hbar v_{\bm k} \partial f_0/\partial \varepsilon_{\bm k}$
Note that here we focus only on the longitudinal current due to the electric field along the $\hat{x}$-direction. Using the velocity for the Dirac Hamiltonian and performing the band index summation, we obtain
\begin{align}
    j_x^{(1)}(t) = -\frac{2e^2 E_0^x}{\hbar^2} \sum_{\bm k} \bigg( \hbar^2 t^2 + \frac{\hbar^4 v^4 k_x^2}{\varepsilon_0^2}\bigg) \pdv{f_0}{\varepsilon_{\bm k}} \frac{e^{-i\omega t}}{1/\tau - i\omega}. 
\end{align}
After the integration over momentum, we get
\begin{align}
    j_x^{(1)}(t) = -\frac{e^2 E_0^x \mu}{2\pi\hbar^2} \bigg( 1 + \frac{2t^2}{v^2} - \frac{\Delta^2}{\mu^2} \bigg)  \frac{e^{-i\omega t}}{1/\tau - i\omega}. 
\end{align}
To simplify the above equation, we consider the low-temperature case, thus first replacing the derivative of the Fermi function with $\delta(\mu - \varepsilon_{\bm k})$ having $\mu$ the chemical potential. Later, we expand the delta function, assuming the tilt term is smaller than the $\varepsilon_0$. Our main focus is to consider the odd-order current contributions which are required for our analysis. In the high-frequency limit, $\omega \tau \gg 1$, we obtain
\begin{align}
    j_x^{(1)}(t) = -\frac{e^2 E_0^x \mu}{2\pi\hbar^2} \bigg( 1 + \frac{2t^2}{v^2} - \frac{\Delta^2}{\mu^2} \bigg)  \frac{i}{\omega}e^{-i\omega t}.
\end{align}
\begin{figure}[ht]
    \centering
    \includegraphics[width=8cm]{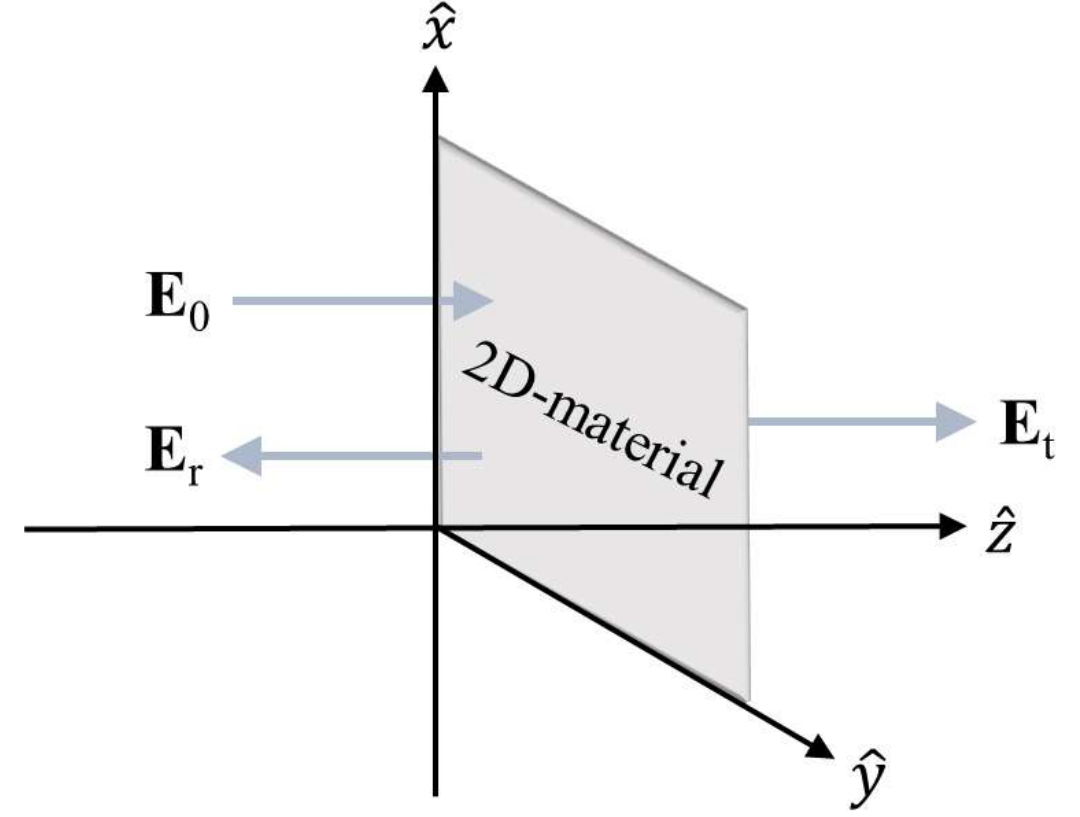}
    \caption{Geometrical picture to interpret the electromagnetic field's boundary conditions. The two-dimensional tilted Dirac material is placed at $z=0$ boundary and is shown in grey color. The symbols $E_0$, $E_r$, and $E_t$ represent the incident, reflected, and transmitted fields respectively.}
    \label{fig:SC}
\end{figure}
From the above equation, we infer that the linear response is directly proportional to the chemical potential which is consistent with the case of graphene ($\Delta =0$ and $t=0$)~\cite{peres_PRB2014}. In addition, the tilt factor adds up the magnitude of the response and the gap reduces the overall value of the linear-order response. Keeping the idea in mind, the third-order current is written in the form
\begin{align} \nonumber
      j_x^{(3)} = &-\frac{e^4}{\hbar^3}(E_0^x)^3 \sum_{\bm k} v_{{\bm k}_x} \frac{\partial^3 f_0({\bm k})}{\partial k_x^3} \\ \nonumber
    & \times \bigg[ \frac{e^{-3i\omega t}}{(1/\tau - i\omega)(1/\tau - i 2\omega)(1/\tau - i 3\omega)} \\[2ex]
    &+  \frac{e^{-i\omega t}}{(1/\tau -i \omega)^2(1/\tau - i 2\omega)} + \frac{2 e^{-i\omega t}}{(1/\tau + i\omega)(1/\tau -i\omega)^2} \bigg].
\end{align}
On further simplifications, the third-order current in the high-frequency limit becomes
\begin{align}\nonumber
      j_x^{(3)} \simeq &  -\frac{3e^4}{4\pi\hbar^2} (E_0^x)^3 \bigg[-\frac{3v^2}{\mu} + \frac{\Delta^2 v^2}{\mu^3} \bigg\{ 1  - \frac{24t^2}{v^2}\bigg\} \bigg]\\[2ex]
      & \times \bigg[ \frac{i}{ 6\omega^3}e^{-3i\omega t}
    -  \frac{i}{2\omega^3} e^{-i\omega t} \bigg].
\end{align}
\section{Optical Bistability}
\label{sec:OB}
Consider that the tilted Dirac sample is located at $z=0$ and the field $E_0e^{ik z}$ is incident on the 2D material as shown in Fig.~\ref{fig:SC}. According to the boundary conditions, the tangential components of the electric and magnetic fields follow \cite{jackson_book}
\begin{align}\nonumber
    E_L &= E_R, \\
    B_L - B_R &= \mu_0 j_x.
    \label{eqn:BC}
\end{align}
\begin{figure*}[ht]
    \centering
    \includegraphics[width=18cm]{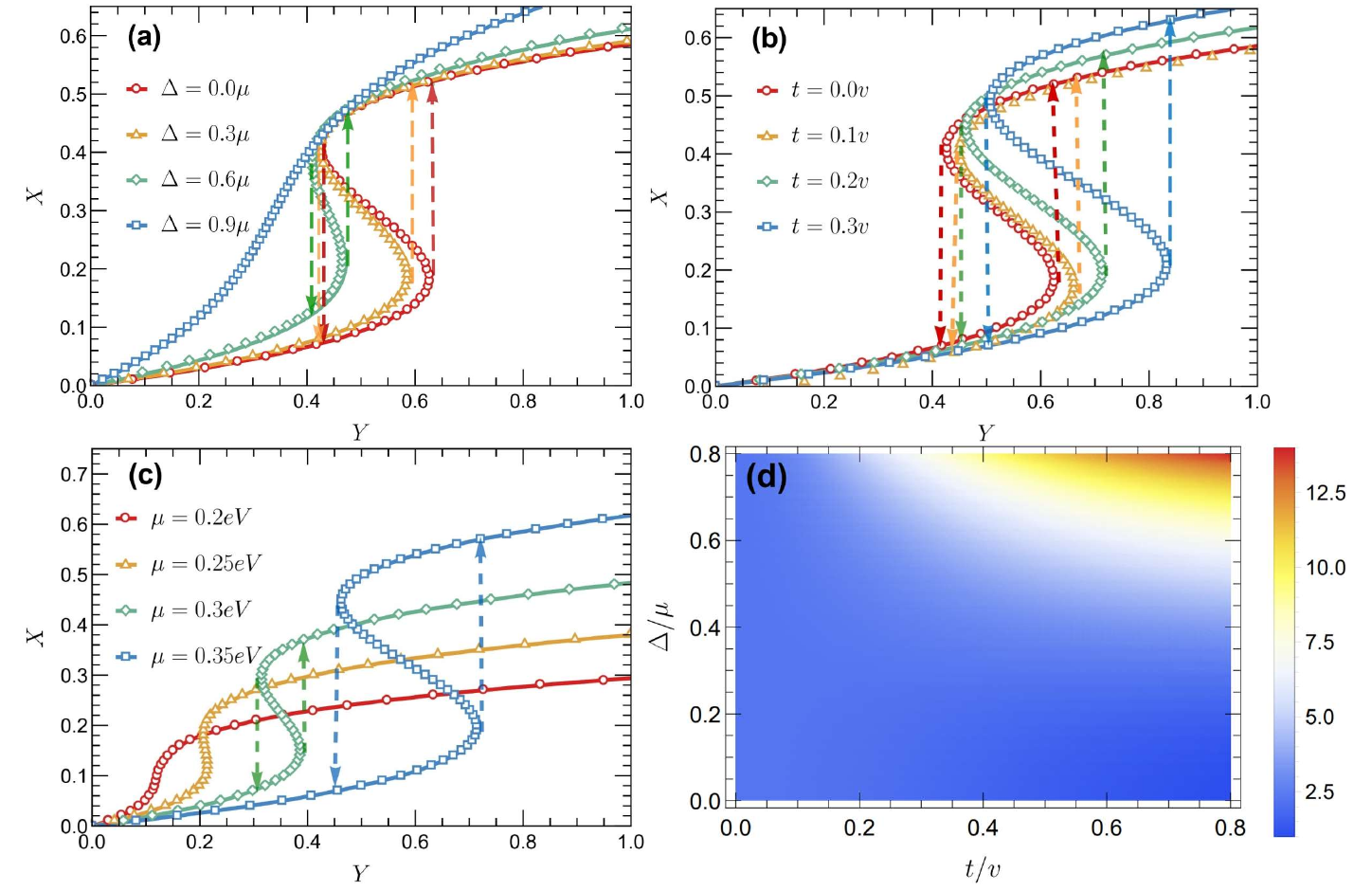}
    \caption{Plot for the Optical bistability variation for a tilted Dirac system. Here, $X$ represents the dimensionless transmitted intensity parameter and $Y$ refers to the incident intensity parameter. Panel (a) shows the variation of the bistability curve at different gap $\Delta$ values and keeps the tilt parameter off. Panel (b) shows the bistability curve at different tilt parameters $t$ and considers the zero gap. (c) optical bistability at different values of the chemical potential, keeping $\Delta = 0$ and $t=0$ (d) gives the density plot of $\lambda$ as a function of normalized gap and tilt parameters. Here, the vertical straight lines with up and down arrows indicate the sudden jumps in the transmitted intensity at the switching upfield $Y_{up}$ and downfield $Y_{down}$.}
    \label{fig:OB}
\end{figure*}

Here $E_{L,R}$ are the electric fields to the left and right side of the sample and are defined as $E_L = E_0 + E_r$, $E_R = E_t$ in which $E_0$, $E_r$, and $E_t$ correspond to the incident, reflected, and transmitted amplitudes of the fields. The magnetic fields $B_{L,R}$ using Maxwell's equations are given by the relations $\omega B_L = k (E_0 - E_r)$ and $\omega B_R = k E_t$ having $\omega$ a frequency of the incident beam and $\mu_0 $ is the permeability of free space. Here we consider the time-dependent electromagnetic fields which oscillate with factor $e^{i\omega t}$. Thus based on the boundary condition, we require the corresponding oscillating current to equate the terms on both sides of Eq.~\eqref{eqn:BC}. Note that we need only those current components proportional to $e^{i\omega t}$. Therefore, the electrical current along the $\hat{x}$-direction in the powers of the electric field is written as
\begin{align}
    j_x &= j_x^{(1,\omega)} + j_x^{(3,\omega)} + \cdots~~.
\end{align}
The current $j_x^{(N,\omega)}$ refers to the component proportional to the $N^{\rm th}$ power of the electric field $E^N$. We keep ourselves only up to third-order because higher-order terms, such as fifth, seventh, etc., do not affect the behavior of the optical bistability. The first and third-order components of the current in response to the external electric field along $x$-direction $E_x$ are defined in the form
\begin{align}
    j_x^{(1,\omega)} &= \sigma_{xx}^{(1)}(\omega) E_0^{x},\\
    j_x^{(3,\omega)} &= \sigma_{xxxx}^{(3)}(\omega,\omega,\omega,-\omega) E_0^{x} |E_0^{x}|^2.
\end{align}
Here quantities $\sigma_{xx}^{(1)}$ and $\sigma_{xxxx}^{(3)}$ represent the first-order (linear) and third-order conductivity tensors. On substituting the longitudinal current in Eqs.~\eqref{eqn:BC} for boundary conditions, we obtain
\begin{align}
    E_0 &= E_t \bigg\{ 1 + \frac{\mu_0 c}{2} (\sigma_{xx}^{(1)} - \sigma_{xxxx}^{(3)} |E_t|^2)  \bigg\}.
\end{align}
This gives the relation between the incident and transmitted fields via the responses. Taking the norm of the above equation while assuming the transmitted field as real, the relation in the compact form reads
\begin{align}
    Y &= X \bigg\{ 1 + \beta (1- \lambda X)^2  \bigg\}.
    \label{eqn:OB}
\end{align}
The above equation is the central equation to analyze the optical bistability~\cite{peres_PRB2014} where we define $X = |E_t|^2$, $Y = |E_0|^2$, and the other parameters as
\begin{align}
    \beta &= \bigg(\frac{\mu_0 c}{2} \sigma_{xx}^{(1)}\bigg)^2;  ~~ \lambda = \bigg(\frac{\sigma_{xxxx}^{(3)}}{\sigma_{xx}^{(1)}}\bigg).
\end{align}
From Eq.~\eqref{eqn:OB}, it follows that for a given value of incident intensity $Y$; there will be more than one real solution for transmitted intensity $X$. However, for $\lambda = 1/X$, Eq.~\eqref{eqn:OB} follows the $Y = X$ relation. This tells that the light does not have any absorption but rather is completely transmitted. However, the emerging latter situation is unrealistic and can not be observed in real problems. For example, in the case of graphene it is shown that there is a small, but finite $2.3\%$ absorbance and $97.7\%$ transmission occur~\cite{Nair_sc2008}. For $\lambda = 0$ or zero third-order response, the strength of the transmitted intensity is reduced by $1/(1+\beta)$ factor. Thus, the actual behavior of transmitted intensity $X$ is controlled by $\beta$ and $\lambda$ parameters, which depend on the strength of the conductivity tensors. 
\subsection{Limiting cases of optical bistability for tilted Dirac system}
For the given tilted Dirac system, the parameters $\beta$ and $\lambda$ using the expressions for the linear and nonlinear responses are
\begin{align} \label{eqn:beta}
    \beta = \frac{\mu_0^2 c^2}{4} \frac{ \mu^2 e^4}{4 \pi^2 \hbar^4 \omega^2} \bigg[1+\frac{2 t^2}{v^2}-\frac{\Delta^2}{\mu^2}\bigg],
\end{align}
\begin{align} \label{eqn:lambda}
    \lambda = \frac{9 e^2 v^2}{4 \mu^2 \omega^2}\frac{1 -\frac{\Delta^2}{3\mu^2}\big\{1-\frac{24 t^2}{v^2}\big\}}{1+\frac{2 t^2}{v^2}-\frac{\Delta^2}{\mu^2}}.
\end{align}
Inserting the Eqs.~\eqref{eqn:beta} and \eqref{eqn:lambda} in Eq.~\eqref{eqn:OB}, we obtain
\begin{align} \label{eqn:YXb}\nonumber
    Y =  & X \bigg\{1 + \frac{\mu_0^2 c^2}{4} \frac{ \mu^2 e^4}{4 \pi^2 \hbar^4 \omega^2} \bigg( 1+\frac{2 t^2}{v^2}-\frac{\Delta^2}{\mu^2}
    - 
    \frac{9 e^2 v^2}{4 \mu^2 \omega^2}\\ 
&\times \bigg[1 - \frac{\Delta^2 }{3\mu^2}\bigg\{1-\frac{24 t^2}{v^2}\bigg\}\bigg]X\bigg)^2\bigg\}.
\end{align}
This is the main equation to understand the optical bistability in terms of tilt and gap parameters in the different limiting regions.\\
\textbf{Case-I:} When there is no gap and finite tilt i.e., $\Delta=0$, $t\neq0$ in the system, Eq.~\eqref{eqn:YXb} takes a form
\begin{align} \label{eqn:YPM}
    Y = X \bigg\{1 + \frac{\mu_0^2 c^2}{4} \frac{ \mu^2 e^4}{4 \pi^2 \hbar^4 \omega^2} \bigg( 1+\frac{2 t^2}{v^2}
    - 
    \frac{9 e^2 v^2}{4 \mu^2 \omega^2} X\bigg)^2\bigg\}.
\end{align}
The above is a cubic equation in $X$ and the roots of the equation can be obtained by setting the discriminant of the equation equal to zero which will give the quadratic equation in $Y$. On solving the quadratic equation, the roots for $Y$ come out to be
\begin{align}
    Y_\pm = \frac{1}{27 \beta \lambda}\bigg[\beta \big(\beta + 9\big) \pm \sqrt{\beta \big(\beta - 3\big)^3}\bigg]
\end{align}
Note that in this case $\beta = \frac{\mu_0^2 c^2}{4} \frac{ \mu^2 e^4}{4 \pi^2 \hbar^4 \omega^2} \bigg[1+\frac{2 t^2}{v^2}\bigg]$. Further, the behavior of the roots depends on the strength of the $\beta$ parameter which affects the occurrence of the optical bistability. Below, we elaborate on these cases.
\begin{enumerate}
    \item For $\beta > 3$, the inequality for the tilt becomes
    \begin{align}
        \frac{t^2}{v^2}> \frac{1}{2}\bigg(\sqrt{\frac{3}{A}}-1\bigg),
    \end{align}
    where $A = \frac{\mu_0^2 c^2}{4} \frac{ \mu^2 e^4}{4 \pi^2 \hbar^4 \omega^2}$. Both the roots for $Y$ remain real. Therefore, under this condition, the optical bistability remains preserved.  
    
    \item 
    For $\beta\leq3$, we have
    \begin{align}
        \frac{t^2}{v^2} \leq \frac{1}{2}\bigg(\sqrt{\frac{3}{A}}-1\bigg).
    \end{align}
    Here, there will be only one real root for $Y$, hence no bistability will occur.
\end{enumerate}
\textbf{Case-II:} When $\Delta\neq 0$ and $t=0$, Eq.~\eqref{eqn:YXb} gives
\begin{align}\nonumber
    Y = &X \bigg\{1 + \frac{\mu_0^2 c^2}{4} \frac{ \mu^2 e^4}{4 \pi^2 \hbar^4 \omega^2}\\
    &\times\bigg( 1-\frac{\Delta^2}{\mu^2}
    - 
    \frac{9 e^2 v^2}{4 \mu^2 \omega^2}\bigg[1 - \frac{\Delta^2 }{3\mu^2}\bigg]X\bigg)^2\bigg\}.
\end{align}
\begin{enumerate}
    \item For $\beta > 3$, the inequality for the gap parameter obtains  
    \begin{align}
    \frac{\Delta^2}{\mu^2}<\bigg(\sqrt{\frac{3}{A}}-1\bigg).
    \end{align}
    Here, the bistability remains intact due to the real roots of $Y$.
\item On taking $\beta\leq3$, we get
    \begin{align}
    \frac{\Delta^2}{\mu^2}\geq \bigg(\sqrt{\frac{3}{A}}-1\bigg).
    \end{align}
    In this case, there will be no bistability.
\end{enumerate}

\section{Results and Discussion}
\label{sec:results}
With the high incident field, it is not sufficient to take the linear contribution of the conductivity to the current, thus required to go beyond the linear regime. Taking the nonlinearity into account, the total conductivity for the tilted Dirac system is expressed as $\sigma = \sigma^{(1)} + \sigma^{(2)}E + \sigma^{(3)} E^2$, where $\sigma^{(i)}$ refers to the $i^{\text{th}}$-order conductivity. However, the optical bistability calculation requires only the odd-order conductivity, as discussed before. Thus, the actual conductivity associated with the factor $e^{i\omega t}$ in the present case is written as $\sigma = \sigma^{(1)} + \sigma^{(3)} |E|^2$. To gain insights into the calculation for the optical bistability, the results for the tilted Dirac system are illustrated in Fig.~\ref{fig:OB}, where the modulus of the transmitted electric field is plotted against the modulus of the incident electric field in Fig.~\ref{fig:OB}(a) and (b).

In Fig.~\ref{fig:OB}(a), we discuss the effect of the gap on the bistability curve. Here, we set the tilt parameter to zero. Notably, the hysteresis loop indicates the optical bistable behavior for $\Delta/\mu < 1$ i.e., the gap is smaller than the chemical potential. It is observed that both lower and upper threshold values vary with the increase of the gap value. Moreover, for a specific value of gap, for instance, $\Delta = 0$ the curve follows the following trend. First, on increasing the input intensity, the transmitted intensity increases. The feature occurs due to the enhancement of the nonlinearity in the tilted Dirac system. When the incident intensity or input value reaches the up threshold value (here $Y_{\rm up} = 0.61$ for $\Delta = 0$), the transmitted intensity or output value $X$ suddenly jumps to a higher value. Further on decreasing the electric field to the down threshold value (here $Y_{\rm down} = 0.41$), the output value jumps to a smaller value. When $\Delta = 0.9 \mu$, the bistable feature vanishes. This indicates the determination of the optical bistability when $\Delta < \mu$ and disappears when the gap approaches the chemical potential value. Thus to observe the optical bistability effect, the chemical potential should lie within the conduction band for a system with the finite gap. 

To study the effect of tilt, we plot the bistability curve in Fig.~\ref{fig:OB}(b) at different tilt values and keeping $\Delta = 0$. We find that the width of the hysteresis loop increases with the tilt. For a specific value $t = 0.3 v$, the maximum width of the threshold occurs at $Y=0.83$. Once the tilt reduces to zero, the width of the threshold decreases, but the optical bistability remains intact. This results from the fact that the $\beta$ factor which relies on the linear response, increases with tilt. However, the tilt does not show significance in the factor $\lambda$, a ratio of nonlinear to linear response. The third-order response shows quartic power law behavior with tilt $\propto t^4/v^4$ and the first-order response shows the quadratic behavior with tilt $t^2/v^2$. This makes the ratio ($\lambda$) proportional to $t^2/v^2$, which is negligible due to the smallness of the tilt term. Notice that we consider the tilt term smaller than the spin-orbit coupling term. This can be visualized in the density plot in Fig.~\ref{fig:OB}(d) for the ratio of nonlinear to linear response as a function of gap and tilt. 

 Furthermore, we observe that the optical bistability can also be tuned via changing the chemical potential even in the absence of the bandgap and tilt in Fig.~\ref{fig:OB}(c). Here, with the increase of the chemical potential the width of the optical bistability loop increases. This happens due to the variation in the level of the absorption. Such variations in the optical properties of the two-dimensional Dirac systems via shifting the chemical potential which can be accessed by doping make them a good candidate for optoelectronic devices.

Based on the numerical estimation, we find that at the chemical potential $\mu = 0.5$ eV, tilt $t = 0.3 v$, 1THz frequency and gap $\Delta = 0$, the roots of $Y$ (Eq.~\eqref{eqn:YPM}) are $Y_+ = Y_{up} = 0.83$ and $Y_- = Y_{down} = 0.45$. This gives the switching upfield (which is the square root of $Y$) as $0.91$ MV/m and the switching downfield as $0.67$ MV/m. However, for $t=0$ case the switching upfield is $0.78$ MV/m, and the switching downfield is $0.64$ MV/m. Therefore, with the finite tilt, the upfield increases by $0.14$ MV/m and downfield by $0.03$ MV/m. However, the values decrease with increasing the gap value. Our results propose that under certain conditions for tilt and gap values the linear and nonlinear responses boost and hence, the optical bistability. In addition, the introduction of tilt provides a new degree of freedom to obtain a desirable optical bistability threshold.

Notice that the tunability of the nonlinear responses, hence the optical bistability can be generated when the current is measured in the direction in which the tilt is considered in the system. On the other hand, the optical bistability via the transverse component of the current due to the field along $y$-direction does not generate any variation with the tilt due to the tilt-independent behavior of the anomalous velocity and the vanishing linear and nonlinear contribution by the product of the band velocites along $x$ and $y$-directions. 

\section{Summary}
\label{sec:summary}
To summarize, we have studied the nonlinear optical bistability phenomenon to examine the effect of tilt and gap. The analysis is performed using the Boltzmann treatment under relaxation time approximation. We found that the width of the hysteresis loop is influenced by the strength of the tilt and gap parameters. Moreover, the switching up and switching down electric field values are increased with tilt and decreased with the Dirac gap. In addition, the switching values mainly depend on the nonlinear response of the system. Specifically, our analysis sheds light on the behavior of nonlinear optical phenomena. Although the study is done at a low THz frequency, the formalism is general and applicable to other materials. Moreover, our study opens new avenues in the direction of optical devices.

\section*{Acknowledgment}
This work is financially supported by the Science and Engineering Research Board-State University Research Excellence under project number SUR/2022/000289.

\bibliography{ref}

\end{document}